\documentclass[conference]{IEEEtran}
\IEEEoverridecommandlockouts
\usepackage[colorlinks,urlcolor=blue,linkcolor=blue,citecolor=blue]{hyperref}
\usepackage[most]{tcolorbox}
\expandafter\def\expandafter\UrlBreaks\expandafter{\UrlBreaks\do\/\do\*\do\-\do\~\do\'\do\"\do\-}
\usepackage{color}
\usepackage{amssymb}
\usepackage{circledsteps}
\usepackage{cite}
\usepackage{hyperref}
\usepackage{amsmath,amssymb,amsfonts}
\usepackage{algorithm}

\usepackage{textcomp}
 \usepackage{array}        
\usepackage{tikz}         
\usepackage{caption}      
\usepackage{mathtools}
\usepackage{algpseudocode}
\usepackage{graphicx}
\usepackage{comment}
\usepackage{url}
\usepackage{textcomp}
\usepackage{multirow}

\begin{document}

\title{When the Code Autopilot Breaks: Why LLMs Falter in Embedded Machine Learning}

\author{\IEEEauthorblockN{Roberto Morabito}
\IEEEauthorblockA{\textit{EURECOM} \\
Biot, France \\
roberto.morabito@eurecom.fr}
\and
\IEEEauthorblockN{Guanghan Wu}
\IEEEauthorblockA{\textit{University of Helsinki} \\
Helsinki, Finland \\
guanghan.wu@helsinki.fi}
}

\maketitle
\begingroup
\renewcommand\thefootnote{}\footnotetext{\noindent\footnotesize This paper has been accepted for publication in Computer (IEEE). Upon publication, the copyright will be transferred to IEEE.}%
\addtocounter{footnote}{-1}
\endgroup

\begin{abstract} Large Language Models (LLMs) are increasingly used to automate software generation in embedded machine learning workflows, yet their outputs often fail silently or behave unpredictably. This article presents an empirical investigation of failure modes in LLM-powered ML pipelines, based on an autopilot framework that orchestrates data preprocessing, model conversion, and on-device inference code generation. We show how prompt format, model behavior, and structural assumptions influence both success rates and failure characteristics, often in ways that standard validation pipelines fail to detect. Our analysis reveals a diverse set of error-prone behaviors, including format-induced misinterpretations and runtime-disruptive code that compiles but breaks downstream. We derive a taxonomy of failure categories and analyze errors across multiple LLMs, highlighting common root causes and systemic fragilities. Though grounded in specific devices, our study reveals broader challenges in LLM-based code generation. We conclude by discussing directions for improving reliability and traceability in LLM-powered embedded ML systems.
\end{abstract}

\maketitle


\section{Introduction}
Automating embedded machine learning (ML) workflows, especially for resource-constrained IoT devices, remains a highly complex challenge \cite{shafique2021tinyml}. From data processing to model deployment, each stage in the embedded ML lifecycle requires domain-specific expertise, toolchain orchestration, and careful hardware alignment. These challenges are amplified in large-scale deployments, where manual coordination becomes impractical. 
While traditional automation frameworks exist \cite{janapa2023edge}, they often address isolated tasks, such as model quantization or deployment, but leave the coordination between stages to human developers. In particular, there is no end-to-end tooling that automates the entire pipeline from raw dataset ingestion to deployable microcontroller code. This lack of integration increases development time and limits scalability across hardware platforms and application domains.

In parallel, recent advances in Large Language Models (LLMs) have opened new opportunities for integrating natural language understanding and code generation into ML workflows \cite{fakhoury2024llm, fan2023large}. Yet, using LLMs as "plug-and-play" components within embedded pipelines is far from trivial. For example, the automation of \textit{deployment sketches}, i.e., the executable code responsible for running the final ML inference on-device, is not an isolated task. It depends critically on the correctness and alignment of earlier pipeline stages (e.g., model conversion – MC), seamless interaction with device I/O, and tight coupling with the hardware-specific constraints of the target platform. These limitations raise a fundamental question: \textit{Can LLMs be relied upon to automate critical stages of embedded ML workflows, and if not, why do they fail?}

\begin{figure*}[!t]
    \centering
    \includegraphics[width=\textwidth]{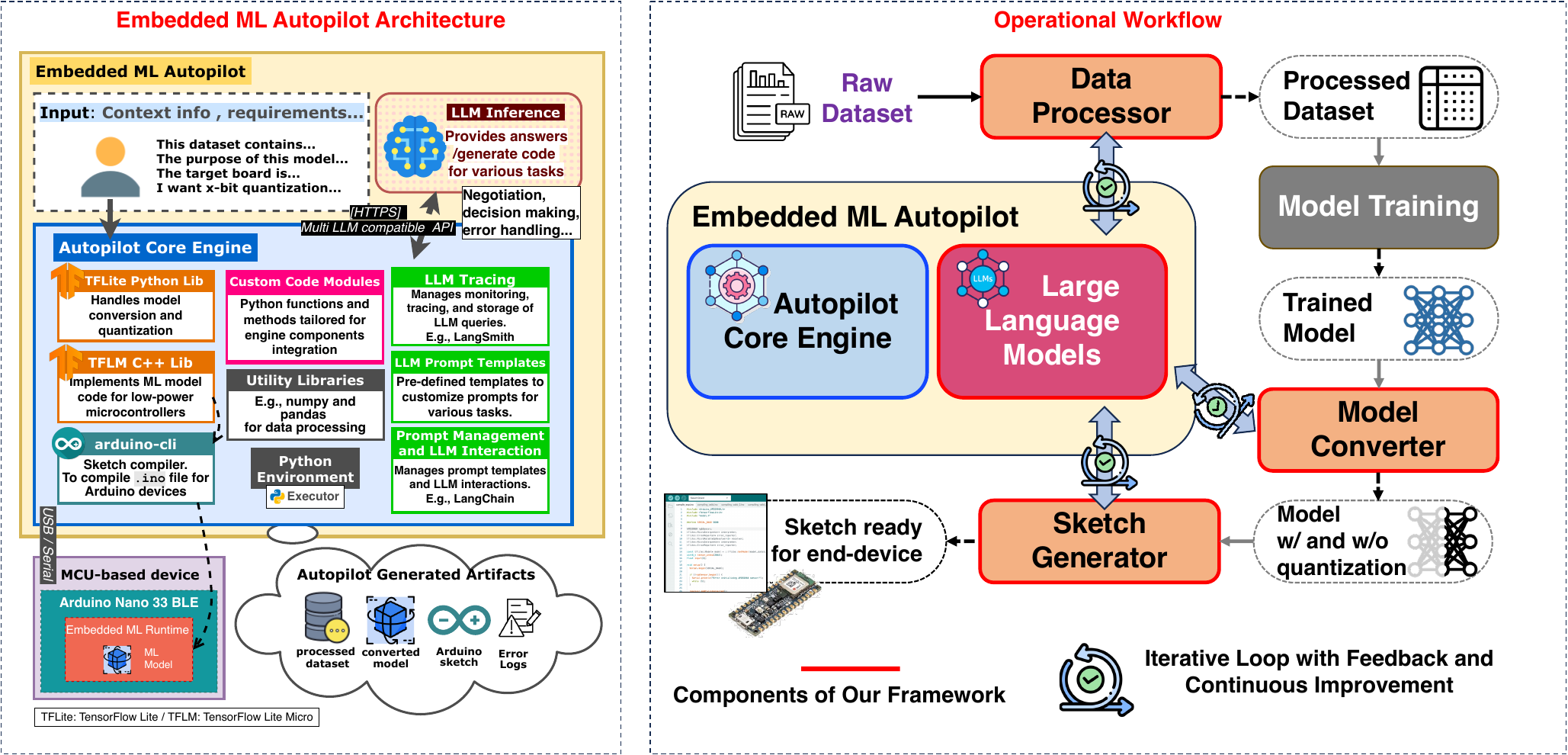}
    \caption{Left: System architecture of the Embedded ML Autopilot, showcasing its internal modules, LLM integration, and runtime environment for managing ML model preparation and deployment. Right: Operational workflow enabled by the Autopilot, outlining the ML lifecycle stages from dataset ingestion to device-ready sketch generation.}
    \label{fig:architecture}
\end{figure*}

In this article, we report on a comprehensive system we built to explore this question. Rather than conducting isolated prompt tests, we developed and evaluated an end-to-end middleware framework that orchestrates LLM interactions across key embedded ML lifecycle stages \cite{wu2025consolidating}. Our system integrates structured prompt design, iterative feedback loops, local validation steps, and tooling integration with embedded ML libraries (e.g., TensorFlow Lite) and constrained IoT devices such as Arduino. This system, which we refer to as the \textit{Embedded ML Autopilot}, was not only designed to reduce human effort, but it also served as a practical lens through which we encountered, firsthand, the limitations and failure points of LLM-powered automation pipelines.

Through a detailed case study and cross-model analysis, we uncover a range of error-prone behaviors, including format-induced misalignment, semantic errors that pass compilation, and unexpected behavioral variations across models. Our findings are synthesized into a taxonomy of failure types derived from an analysis of over a thousand log-traced errors collected across multiple LLMs. We show how prompt structure, decoding strategy, and model family influence both success rates and failure patterns, and we identify a specific lifecycle stage, sketch generation (SG), as disproportionately fragile.

While combining LLM reasoning with embedded computing systems constraints, this work aims to contribute with practical insights, actionable diagnostics, and forward-looking strategies for building failure-aware AI systems. We argue that reliable automation will require not just better models, but also architectures that can anticipate, detect, and respond to LLM-induced faults \cite{zhang2022towards}, which is an essential step toward scalable, intelligent automation in the embedded AI domain.

\section{Background and Context}

Automating embedded ML pipelines requires bridging high-level model reasoning with the realities of low-level hardware constraints. In this context, our work builds on a comprehensive system architecture, the \textit{Embedded ML Autopilot}, designed to orchestrate interactions between LLMs and the stages of the embedded ML workflow. The Autopilot framework enables not only the generation of deployment artifacts (such as microcontroller-ready code) but also serves as an experimental infrastructure for observing and diagnosing model behavior across stages.

\subsection{System Overview: The Embedded ML Autopilot}
Figure \ref{fig:architecture} outlines the two central components of our system: the \textbf{Autopilot Core Engine} and the \textbf{Operational Workflow}.

The left panel presents the architectural components of the Embedded ML Autopilot. At the heart of the system is the Core Engine, which integrates a range of task-specific modules: prompt templates, utility libraries, API handling logic, and dedicated components for LLM interaction and validation. These modules coordinate actions such as MC, quantization, and SG. Notably, the engine supports different LLM families and backends and is designed to operate with limited or no internet connectivity, which is highly important for private or local deployments.

The right panel illustrates the end-to-end data flow. The workflow begins with raw dataset ingestion and proceeds through a sequence of transformation steps, including data preprocessing (DP), model training and conversion, and deployment SG. At each step, the Autopilot invokes the LLM to generate intermediate outputs, which are validated and refined through feedback loops. This design allows the framework to simulate real-world embedded ML development conditions, where generated code must align with hardware constraints, available libraries, and execution requirements.

\begin{figure*}[!t]
    \centering
    \includegraphics[width=\textwidth]{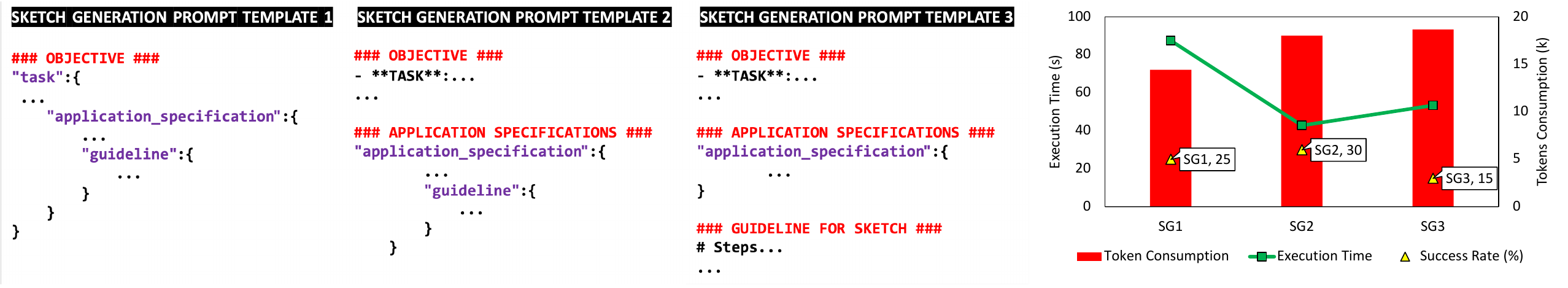}
    \caption{Comparison of three prompt templates (SG1, SG2, SG3) used for SG. The results highlight trade-offs between token usage, execution latency, and success rate. SG2 demonstrates the highest success rate, while SG3 minimizes token consumption.}
    \label{fig:sketch_comparison}
\end{figure*}

\subsection{Scope and Focus of This Article}
A full architectural description of the Embedded ML Autopilot is provided in earlier work \cite{wu2025consolidating}. Here, we focus on the behavioral patterns and failure mechanisms observed during its operation. Despite our goal of a fully automated framework for embedded ML, hands-on experience revealed persistent limitations—even with advanced prompt engineering, tailored integration, and extensive iteration. These findings stem from months of engineering, testing, and refinement, exposing a gap between the promise of LLM-driven automation and its real-world reliability.

To study these gaps, the Autopilot logs each interaction, including prompt structure, decoding outcomes, LLM responses, and downstream errors. These logs underpin our failure analysis. Unlike prompt-centric benchmarks, the framework observes full lifecycle behavior, from data ingestion to device-ready code, revealing structural fragility, inconsistencies, and stage-specific failures.

This article examines one demanding scenario: automated generation of executable code (sketch) for a microcontroller-based vision application. The setup involves an ML model on a resource-constrained device (e.g., Arduino), integration with a color sensor, and execution of inference via deployment sketches. We chose this case as a stress test, since it exercises all stages—data handling, model conversion (MC), on-device inference, and hardware-specific code generation—providing a rich space to observe where LLM automation breaks down. While DP and MC generally succeeded with only minor and recoverable issues (e.g., schema mismatches or quantization shape errors), the SG stage showed the lowest reliability, often below 40\% across models and settings. This stage displayed especially fragile behavior across prompt formats and model outputs, making it the focus of our analysis.

\section{Characterizing Failures in Sketch Generation}

We now focus on the SG stage, where we observed the most persistent reliability issues. Rather than concentrating solely on explicit failure states such as syntax errors or missing includes, we take a broader view: examining subtle behavioral inconsistencies, formatting fragility, semantic gaps, and runtime-disrupting edge cases. We highlight how seemingly minor prompt variations can dramatically affect outcomes, how different model families exhibit divergent generation tendencies, and how even well-formed code can silently fail during deployment. These findings lay the foundation for the formal taxonomy introduced in the next section and inform our recommendations for model selection, prompt design, and failure-aware orchestration.

Before proceeding, we define a “successful” SG as one that compiles and executes correctly on the target device. Specifically, the generated \texttt{.ino} file must \emph{(i)} compile, \emph{(ii)} run without runtime errors, and \emph{(iii)} perform the expected behavior, including sensor interaction and ML inference. Sketches that compiled but produced no response, invoked unrelated routines, or skipped inference were considered failures. For validation, we used a vision classification task with an official Arduino reference, verifying that generated sketches matched its structure and logic, particularly sensor I/O handling and inference calls. This ensured our evaluation captured both structural validity and functional reliability.

\subsection{Prompt Structure Sensitivity: A Source of Hidden Fragilities} 

While SG consistently emerged as the most failure-prone stage, a deeper investigation revealed that prompt structure, particularly nested JSON-like objects, significantly influenced LLM behavior. To evaluate this effect, we compared three formatting variants (SG1, SG2, SG3), each differing in how task, specification, and guideline components were embedded.

Regardless of formatting, the prompt content was consistent across experiments \cite{wu2025consolidating}. Each template combined: \emph{(i)} a context block defining the LLM’s assumed TinyML expertise, \emph{(ii)} task-specific objectives (e.g., generating specifications or Arduino sketches), \emph{(iii)} explicit output-formatting rules (single code block, JSON, or C++), \emph{(iv)} error-handling and refinement instructions, and \emph{(v)} a guideline schema for sketch steps (e.g., library inclusion, memory setup, inference invocation). Templates were iteratively refined, informed by prior LLM-for-code studies and pilot runs.

As shown in Figure \ref{fig:sketch_comparison}, SG1 nests the entire prompt (task and guidelines) in JSON, SG2 moves the task outside while keeping specifications and guidelines in JSON, and SG3 flattens further by placing the guidelines in Markdown as a separate block.

\begin{figure*}[!t]
    \centering
    \includegraphics[width=\textwidth]{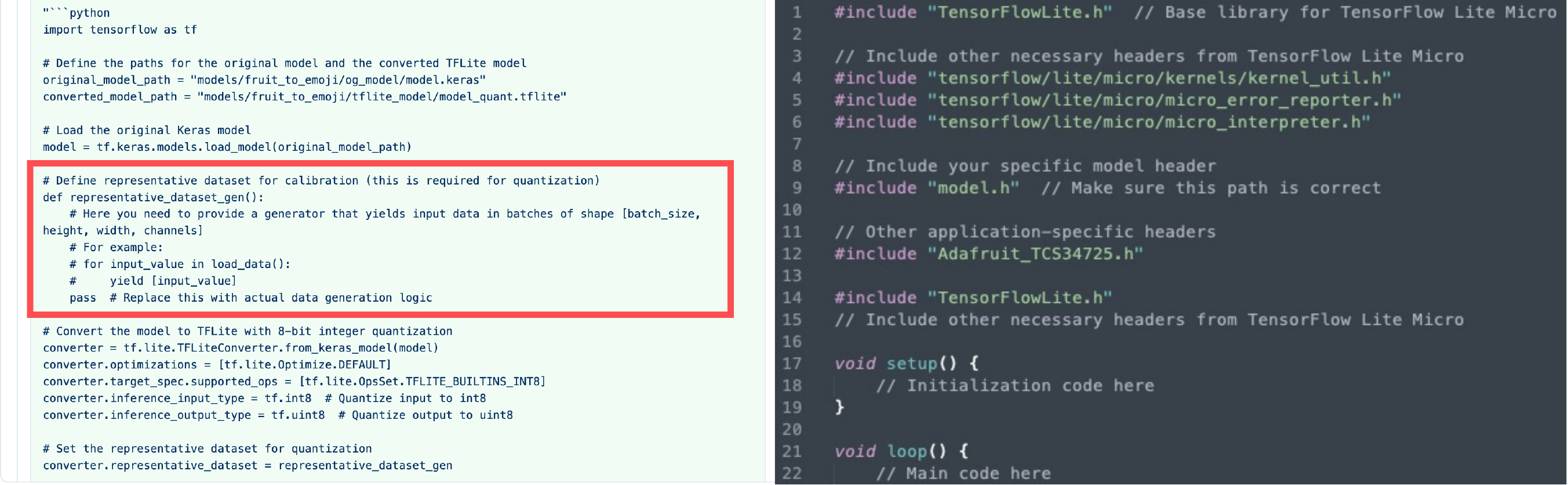}
    \caption{Left: Example of multiple candidate completions, where the LLM embeds alternative code paths inside the explanation, introducing ambiguity and requiring manual disambiguation. Right: Example of a compilable but incomplete sketch, where structurally valid code omits the core inference logic, resulting in silent functional failure.}
    \label{fig:slms}
\end{figure*}

The right side of Figure \ref{fig:sketch_comparison} shows trade-offs across the three prompt templates in token use, latency, and success rate. SG2 achieved the best success (30\%) with moderate tokens and the shortest latency. SG1 performed slightly worse (25\%), with the longest latency despite being most token-efficient. SG3 had the lowest success (15\%), consumed the most tokens, and showed high latency, indicating inefficiency in both reliability and resource use.

These results suggest that LLM performance can degrade not only from ambiguous content but also from structural presentation \cite{errica2024did}, even when semantically equivalent. They further highlight a non-monotonic relationship between resource use and success: more detailed prompts do not necessarily yield better outcomes.

We interpret this as format-induced misalignment, where the LLM’s parsing strategies or token heuristics fail to resolve nested schemas or overloaded formats \cite{tian2023chatgptultimateprogrammingassistant}.

These findings have several practical implications:
\begin{itemize}
    \item[\textbf{(i)}] Prompt structures must be carefully engineered, not only for correctness but for LLM readability under constrained conditions.
    \item[\textbf{(ii)}] Token budget alone is an unreliable predictor of task success, suggesting that iterative retries may only exhaust the cost without gain.
    \item[\textbf{(iii)}] Failure modes linked to structure are often silent, meaning they do not manifest in error messages, making them harder to detect without systematic testing.
\end{itemize}

This prompt structure sensitivity introduces a new class of AI failures—those rooted in format misinterpretation rather than knowledge gaps or logic errors. In the broader context of deploying LLMs in IoT or embedded workflows, this insight calls for integrating prompt testing frameworks \cite{wang2024prompt}, adopting explicit interface contracts, and enforcing consistency checks between stages.

\subsection{Emergent Patterns from Open Models: What They Do (and Don’t) Get Right}

To assess whether SG failures in GPT-4o are unique to closed-source LLMs, we replicated key lifecycle steps using open-source models integrated into our framework. These included instruction-following and code-oriented LLMs, run locally without \textit{over-the-Internet} API calls. Our goal was not to benchmark performance, but to test whether structural and format-related failures generalize across model types.

The results confirmed our hypothesis: open-source models showed the same failure classes, often with additional parsing breakdowns.

To support this analysis, we evaluated several state-of-the-art open-source LLMs—Phi-4, Llama3.1, Qwen2.5-Coder, Deepseek-R1, and Codestral—under the same DP, MC, and SG lifecycle used for GPT-based models. All models were tested under provider defaults for decoding parameters (e.g., \textit{temperature}, \textit{top-p}) to ensure comparability and reproducibility. We note that decoding stochasticity can itself influence the failure surface, an aspect we analyze in extended experiments.

Reliability declined sharply, most prominently in SG, where success fell to 0\% in all but one case. Only Codestral 22B reached $\approx$11\% in SG, while others remained near zero. Even in DP and MC, performance lagged GPT-4o, which routinely exceeded 90\%. For example, Phi-4 14B reached 50\% in DP and 97\% in MC, but only 3\% in SG.

These results highlight that model capability strongly influences tasks requiring multi-step reasoning and structurally valid output \cite{wu2024mindmap}. Yet the recurrence of similar error types across model families suggests failures arise not only from model limitations but also from deeper vulnerabilities in the current LLM-based paradigm, such as weak format enforcement, lack of semantic validation, and prompt fragility \cite{yang2024morepair}.

We highlight two failure patterns that were especially common:
\\
\Circled{Case 1} \textbf{Unparsable or mixed-format output:} Some models produced outputs with interleaved explanation and code, or alternated between multiple code blocks without presenting a complete executable sketch. This caused our automated parsing logic to fail.
\\
\Circled{Case 2} \textbf{Multiple candidate completions without structure:} In several runs, the model proposed alternative snippets or variants, asking the user to "\textit{choose the appropriate one}". While this may be helpful in interactive settings, it introduces ambiguity in automation pipelines. Figure \ref{fig:slms}-left illustrates such a scenario, where multiple solutions are embedded inside the explanation text, requiring human disambiguation.

These examples reinforce that prompt sensitivity \cite{errica2024did} and validation blind spots are not exclusive to proprietary LLMs. Even with full model transparency and local deployment, the lack of enforced output structure and the absence of end-to-end behavioral validation can lead to silent or delayed failures. Moreover, format inconsistencies—such as Markdown misalignment, missing triple backticks, or ambiguous nesting—proved equally disruptive in open-source settings.

\begin{tcolorbox}[
    breakable,
    colback=green!5!white, 
    colframe=green!75!black, 
    title=Mitigation Strategy, 
    fonttitle=\small, 
    fontupper=\small,
    sharp corners, 
    boxrule=0.5pt, 
    left=2pt, right=2pt, top=2pt, bottom=2pt,
    before skip=6pt, after skip=2pt
]
From an engineering standpoint, these findings suggest the need for model-agnostic validation layers, such as: \emph{(i)} output format checkers (e.g., complete sketch in a single block), \emph{(ii)} structural linters for code completeness, and \emph{(iii)} post-generation semantic validators (e.g., checking if inference function is called).
\end{tcolorbox}

\subsection{When Code Compiles but Breaks: Semantic Gaps and Pipeline Disruptions}

In addition to structural issues in prompt formatting and code block generation, our evaluation uncovered more subtle failures. These do not stem from prompt misunderstanding but from semantic mismatches or pipeline-breaking behaviors, even when outputs pass syntactic checks. Such issues relate to the phenomenon of \textit{patch overfitting}, where generated patches pass test cases yet fail to address the root defect, producing semantically incorrect solutions that undermine reliability \cite{petke2024patch}. These failures are particularly concerning in automation pipelines, where validation often relies on compilation, simple I/O checks, or success logs. The following examples illustrate why workflows must go beyond correctness checks and include semantic and contextual validation.
\\
\Circled{Case 1} \textbf{False Positives from Compilable but Incomplete Sketches:} Some SGs produced compilable \texttt{.ino} files that passed structural checks yet failed when deployed. In one case, the sketch invoked a color sensor routine instead of running ML inference. The code was structurally valid, included headers, and compiled without error—but contained no logic for the trained model or input pipeline. This highlights a critical risk: compilation \textbf{$\neq$} functional correctness \cite{chen2024challenges}. Automated workflows may validate compilation and function calls, yet LLM-generated code can silently skip core logic, yielding non-functional deployments. Figure \ref{fig:slms}-right shows such a case, where sensor logic is missing or replaced with placeholder code. 

\begin{tcolorbox}[
    breakable,
    colback=green!5!white, 
    colframe=green!75!black, 
    title=Mitigation Strategy, 
    fonttitle=\small, 
    fontupper=\small,
    sharp corners, 
    boxrule=0.5pt, 
    left=2pt, right=2pt, top=2pt, bottom=2pt,
    before skip=6pt, after skip=2pt
]
LLM-based SG tools require behavioral validation layers, including: \emph{(i)} minimal I/O tracing or signal path verification during deployment, \emph{(ii)} targeted unit testing of generated functions, or \emph{(iii)} testbed-based validation (e.g., observing serial outputs in response to simulated inputs).

\end{tcolorbox}

\noindent\Circled{Case 2} \textbf{Silent Breakage from Incomplete Data Processing:} In the DP stage, we observed scripts that loaded and analyzed a dataset but failed to save the transformed output—despite printing a success message and triggering the next step. In some cases, the script reported a new file path (via JSON) that was never written. This produced false positives in stepwise validation: the system assumed the transformation succeeded, but the next component failed when loading the nonexistent file. Such failures stem from semantic hallucination in logging and formatting, where the LLM generates plausible I/O structures without enforcing file consistency. In long workflows, these inconsistencies propagate silently and are difficult to trace without audit logging or strict output-path checks.
\begin{tcolorbox}[
    breakable,
    colback=green!5!white, 
    colframe=green!75!black, 
    title=Mitigation Strategy, 
    fonttitle=\small, 
    fontupper=\small,
    sharp corners, 
    boxrule=0.5pt, 
    left=2pt, right=2pt, top=2pt, bottom=2pt,
    before skip=6pt, after skip=2pt
]
We emphasize the need for: \emph{(i)} runtime existence checks of output artifacts, \emph{(ii)} semantic consistency verification across stages, and \emph{(iii)} potentially adopting output schema templates to align model-generated paths with expected file structure.
\end{tcolorbox}

\section{Taxonomy of Failure Categories} 

\begin{figure*}[!t]
    \centering
    \includegraphics[width=\textwidth]{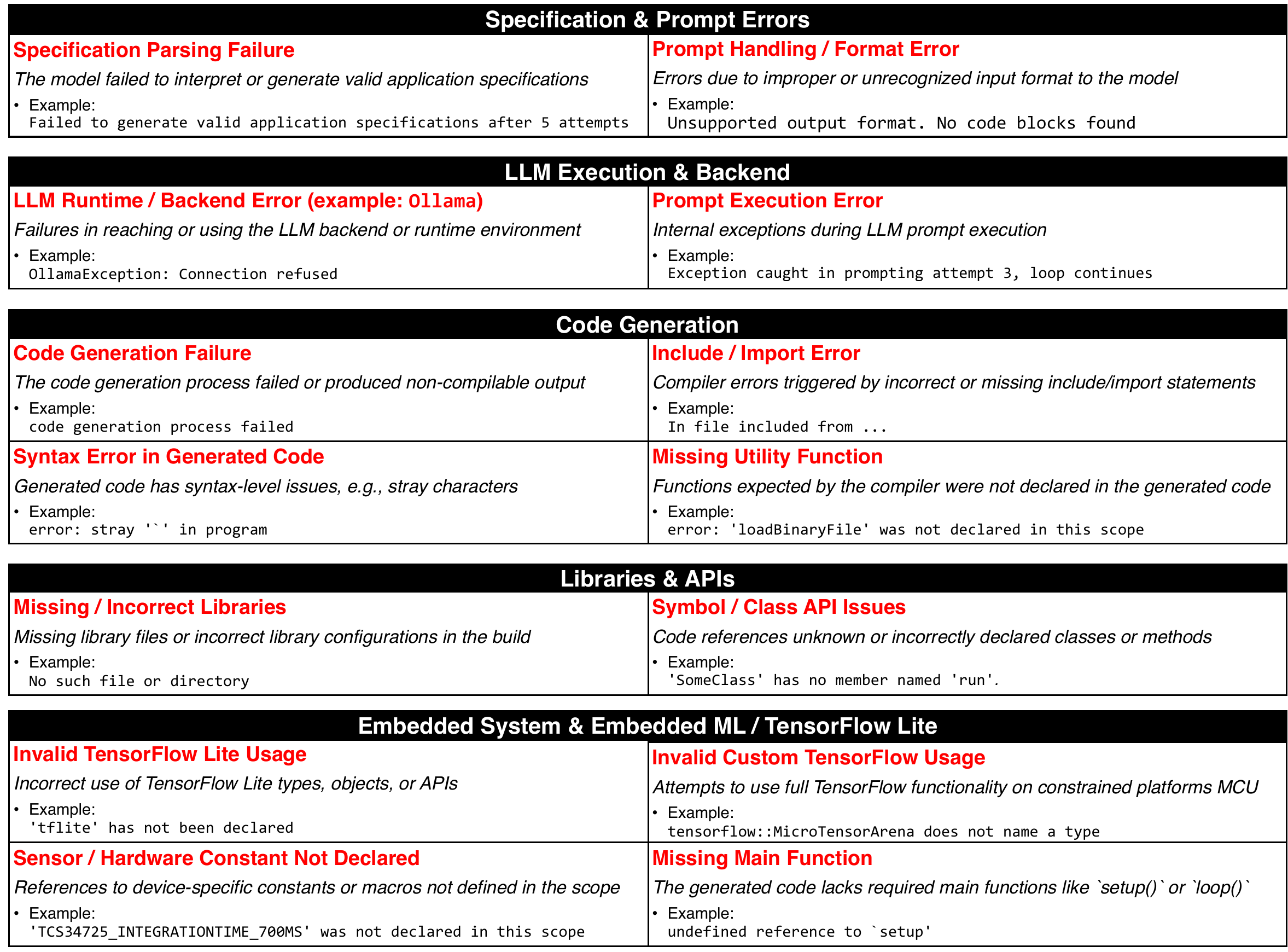}
    \caption{Taxonomy of LLM-driven code generation errors, with descriptions and representative real-world examples.}
    \label{fig:failures_summary}
\end{figure*}

We broaden the view with a taxonomy of failures, derived from systematic log analysis across proprietary and open-source LLMs.

Observed errors reflected diverse causes extending beyond prompt design or token limits \cite{tambon2025bugs}. Analyzing over two hundred failed SG attempts per LLM, we organized them into a multi-layered taxonomy based on where and why generation failed.

The grouped categories, shown in Figure \ref{fig:failures_summary}, span from shallow formatting errors and missing libraries to deeper semantic and API-level faults. While some (e.g., missing \texttt{setup()} functions or malformed headers) are recoverable via retries or static checks, others (e.g., incorrect TensorFlow API use or symbol mismatches) reveal fundamental gaps in LLM reasoning and code alignment.

Several categories also capture failures that compile but break only at deployment or runtime—such as semantic misalignments, placeholder logic, or undeclared sensor constants. These are especially problematic in automated pipelines, silently evading early validation. By formalizing this taxonomy, we move from isolated bugs to systemic design flaws in prompt-driven workflows. Each category reflects not just symptoms, but assumptions about what the model does and does not understand.

\subsection{Error Profiling Across Language Models}

While the taxonomy provides a structured view of \textit{what types of errors occur and why}, it does not reveal \textit{how often} these failures manifest across models, nor whether some errors are systemic or model-specific. To bridge this gap, we complement the taxonomy with an error profiling study across representative LLMs.

\begin{figure*}[!t]
    \centering
    \includegraphics[width=\textwidth]{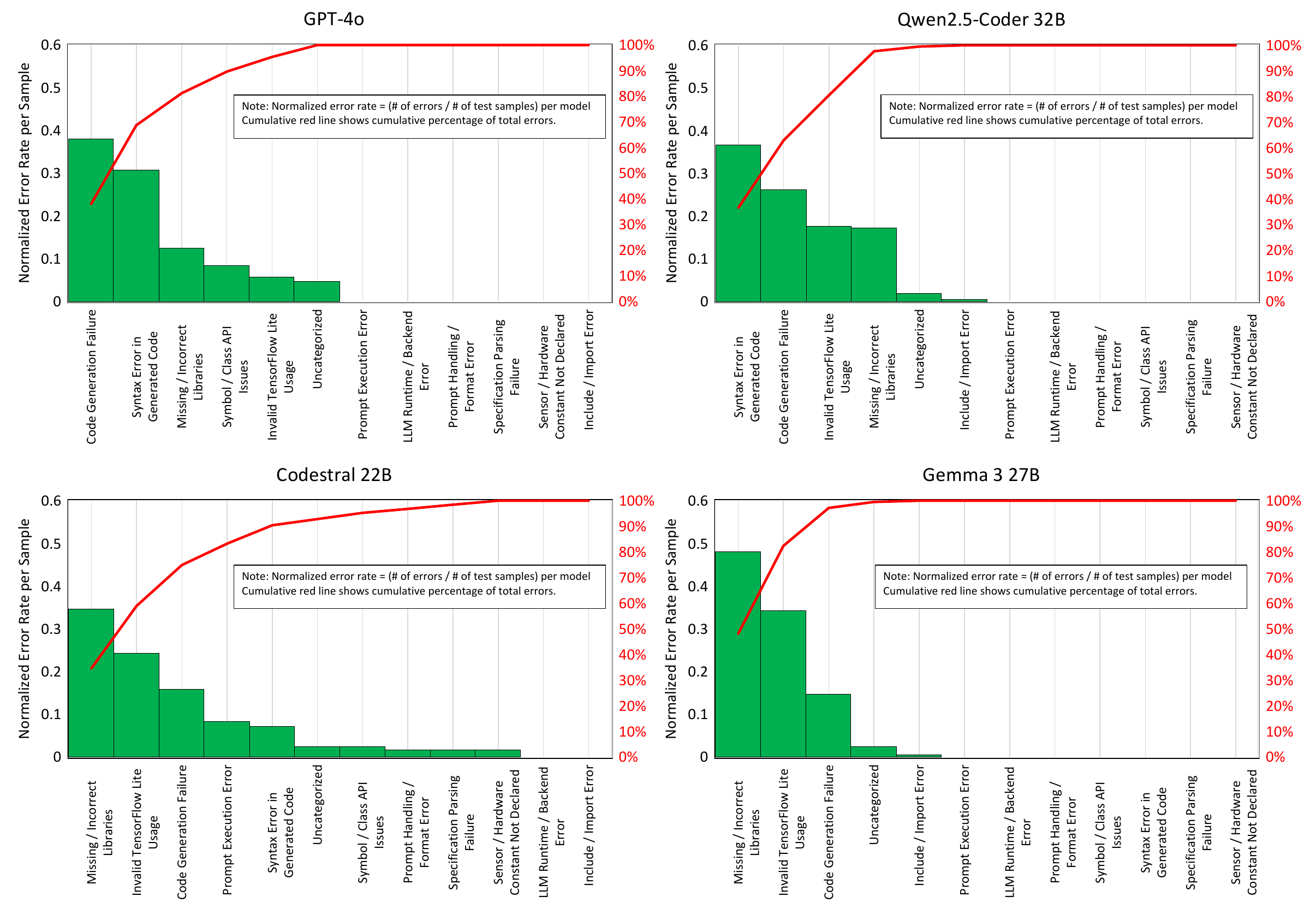}
    \caption{Distribution of error types across four LLMs evaluated in our system. The bars represent the normalized error rate per test sample for each error category, while the red line shows the cumulative percentage of total errors.}
    \label{fig:failures}
\end{figure*}

Figure \ref{fig:failures} summarizes the normalized error rate per failure category for four representative LLMs: GPT-4o (\texttt{gpt-4o-2024-08-06}), Qwen2.5-Coder 32B, Codestral 22B, and Gemma 3 27B.

From this analysis, several key insights emerge. \textit{Code Generation Failure} and \textit{Syntax Errors} dominate across all models, especially GPT-4o and Qwen2.5-Coder. This suggests a recurring inability to produce fully compilable code when following structured multi-step prompts. \textit{TensorFlow Lite misuse}, such as incorrect type references or attempts to invoke functions unsupported on microcontrollers, was especially prominent in open-source models like Codestral and Gemma. This reflects a lack of model grounding in hardware-specific constraints. \textit{Missing or incorrect library usage}, including absent \texttt{\#include} or undeclared utility functions, accounted for a large share of failures, particularly in the open-source models. \textit{Prompt execution and runtime exceptions} were far more common in open-source models, possibly due to weaker instruction following or model misalignment with the JSON-like prompt structure used in the system. Some models (notably GPT-4o) exhibited a wider distribution of low-frequency errors (e.g., undeclared sensor constants, API symbol mismatches), pointing to higher variability in code generation attempts.

\begin{tcolorbox}[
    breakable,
    colback=blue!5!white, 
    colframe=blue!75!black, 
    title=Main Takeaway, 
    fonttitle=\small, 
    fontupper=\small,
    sharp corners, 
    boxrule=0.5pt, 
    left=2pt, right=2pt, top=2pt, bottom=2pt,
    before skip=6pt, after skip=2pt
]
This empirical profiling reinforces the notion that certain failure types are deeply systemic, while others are amplified by specific model weaknesses, such as the lack of embedded ML domain grounding or inconsistent formatting behavior.

\end{tcolorbox}

\begin{tcolorbox}[
    breakable,
    colback=green!5!white, 
    colframe=green!75!black, 
    title=Mitigation Strategy, 
    fonttitle=\small, 
    fontupper=\small,
    sharp corners, 
    boxrule=0.5pt, 
    left=2pt, right=2pt, top=2pt, bottom=2pt,
    before skip=6pt, after skip=2pt
]
Being able to characterize the relative prominence of failure categories per model, we can prioritize mitigation strategies that are both model-agnostic (e.g., checking for syntax or missing includes) and model-specific (e.g., preventing TensorFlow misuse on MCUs). These insights also inform the need for failure-aware orchestration systems, where the choice of model is influenced by its known error tendencies for a given task type.
\end{tcolorbox}

\section{Further Recommendations for Failure-Aware System Design}
Throughout this paper, we have highlighted key takeaways and mitigation strategies drawn from our analysis. While there exists a vast landscape of potential directions for improving LLM robustness, our focus here is on recommendations that align with the core layers of our framework: validation and verification mechanisms, and orchestration strategies centered around prompts and interaction formats. We now turn to these areas, looking ahead to how failure-aware design can be extended and strengthened in future LLM-driven embedded ML automation.

Our experience suggested that beyond ad hoc retries or structural checks, these systems can benefit from embedding explicit I/O tracing, semantic verifiers, and output-structure enforcement within the execution layer. However, while guardrails and validators in prompts can drastically increase success rates, with smaller LLMs we observed that such mechanisms were not sufficient to prevent critical failures, highlighting the need for finer-grained validation strategies. 

Pairing code generation with formal verification methods offers one path \cite{sevenhuijsen2025vecogen}. At the same time, orchestration must cope with the rising complexity of prompt design in embedded ML workflows, where the same autopilot pipeline may target heterogeneous devices, data, and applications. Prompts increasingly combine dynamic components, multi-role interactions, structured data, and diverse output formats, making ad hoc prompt templates fragile \cite{sahoo2024systematic}. Modular frameworks—supporting schema constraints, role awareness, and context injection—can help mitigate this fragility \cite{zhang2025prompt}, but pipelines must also adapt to the provider-specific interaction formats that different LLMs (e.g., Ollama, HuggingFace models) impose in their protocols. 

Building systems that seamlessly align external prompts with these model-specific formats is non-trivial but essential, since different LLMs may excel or fail in different pipeline stages or tasks. Here, neuro-symbolic reasoning can represent a natural complement. For example, symbolic-logic coupling strategies have demonstrated how declarative programming language backends can enforce correctness and stability in multi-step reasoning, suggesting that similar approaches could strengthen prompt adaptation and verification in embedded ML pipelines \cite{di2025lorp}.

Taken together, these directions call for pipeline “autopilots” that merge robust validation, flexible orchestration, and adaptive model integration. Unifying tracing, verification, modular prompt management, and model-aware adaptation can help future systems reach new levels of reliability, scalability, and resilience across heterogeneous devices, tasks, and execution environments.

\section{Conclusion}

We presented an empirical study of failure modes in LLM-driven automation of embedded ML pipelines. Our analysis uncovered recurring patterns of structural, semantic, and behavioral fragility, many of which remain silent or hard to trace. While our work is grounded in a specific set of embedded devices and applications, the failure modes we identified reflect broader challenges intrinsic to LLM-based code generation. We see the presented taxonomy and our recommendations as a starting point for strengthening the intersection of embedded ML and generative AI, where resource constraints amplify the risks of unreliable code generation and underscore the need for failure-aware systems design.

\section{*Note on Reproducibility}

To support transparency and enable further exploration, we have released the SG logs and analysis scripts used in this study via a public GitHub repository\footnote{\url{https://github.com/robertmora/embeddedML-autopilot-failures}}. The provided resources allow reproduction of key results and offer flexibility for extending the failure analysis across decoding settings, models, or error types.

\bibliographystyle{IEEEtran}
\bibliography{References}

\begin{IEEEbiographynophoto}{Roberto Morabito} (Member, IEEE) is an Assistant Professor in the Communication Systems Department at EURECOM, France. His research focuses on networked AI systems, with particular attention to AI service provisioning and lifecycle management under computing and networking constraints. He earned his PhD from Aalto University and has held positions at Princeton University and Ericsson Research. He is also Docent in Computer Science at the University of Helsinki, Finland (roberto.morabito@eurecom.fr).
\end{IEEEbiographynophoto}
\begin{IEEEbiographynophoto}{Guanghan Wu}(Student Member, IEEE) is a Research Assistant in the Department of Computer Science at the University of Helsinki, Finland. He earned his Master’s degree in Computer Science from the same university in 2024 (guanghan.wu@helsinki.fi).
\end{IEEEbiographynophoto}
\vfill
\end{document}